\documentstyle[12pt]{article}
\begin{document}

\title{Test of the Quantum Chaoticity Criterion for
       Diamagnetic Kepler Problem}
\author{V.E. Bunakov, I.B. Ivanov, R.B. Panin
\footnote{e-mails: Vadim.Bunakov@pobox.spbu.ru, Ivan.Ivanov@thd.pnpi.spb.ru}
\\Petersburg Nuclear Physics Institut, 188350, Gatchina, Russia}

\date{}
\maketitle
\begin{center}
{\bf Abstract:}
\end{center}
{\small
The earlier suggested criterion of quantum chaoticity, borrowed from
the nuclear compound resonance theory, is used in the analysis of the
quantum diamagnetic Kepler problem (the spinless charged particle motion
in the Coulomb and homogenious magnetic fields).
}

\section{Introduction}

It is believed for decades that the main feature of the classically
chaotic system is the instability of its trajectories to minor variations
of the initial conditions. Since the concept of the trajectory in phase space
does not apply in quantum mechanics, the possibility of quantum chaos is still
opened to discussion. It was suggested [1] to look for the "quantum signatures
of classical chaos". The only more or less generally accepted "signature"
found up to now is the Wigner level repulsion. In this paper we proceed with
the illustration of our alternative suggestions [2-5]
concerning the definition of chaos for the Hamiltonian quantum (and classical)
systems.

     According to Liouville-Arnold theorem in classical mechanics,
the Hamiltonian system with $N$ degrees of freedom is regular if it has
$M=N$ independent global integrals of motion. If the number $M$ of global
integrals becomes less than $N$, the system becomes chaotic. The well
known Noether's theorem connects the existence of global integrals of
the system with the symmetries of its Hamiltonian. According to this theorem,
breaking the symmetry of the initially regular system decreases the number
of its independent global integrals of motion. Thus the system becomes
chaotic {\bf only} in the case of such a symmetry-breaking which makes the
number $M$ of global integrals {\bf less} than $N$.

Our first (and major) suggestion is to generalize this definition of
chaoticity for the case of quantum systems. Since the concept of
symmetry (unlike the trajectory) is universal for both classical and quantum
mechanics, this
generalization seems to be quite straightforward - one should simply
substitute
the integrals of motion by the corresponding 'good' quantum numbers, resulting
from the symmetries of quantum Hamiltonian. This approach immediately
allows to treat the only generally accepted signature of quantum chaos -
Wigner's level repulsion - as a signature of symmetry-breaking leading to
chaos. Indeed, the general property of the highly symmetrical regular quantum
system is the high degeneracy of its eigenstates. The immediate consequence
of perturbation breaking the original symmetry is the removal of this
degeneracy (in other words, Wigner's level repulsion). It is worthwhile to
remind that Wigner's level repulsion was first observed for the resonance
states of compound nucleus, whose only good quantum numbers are their energy
and spin.  This property comes from the fact that the symmetries of the
nuclear mean field are destroyed by the pair-wise "residual" interactions
[2--5].

Our second (rather technical) suggestion is to use the concept of spreading
width $\Gamma_{spr}$ (and the related criterion $\ae$) as a sensitive measure
of symmetry-breaking of the Hamiltonian $H_0$ caused by the perturbation
$ V$. Indeed, consider a Hamiltonian $H$ of the non-integrable
system as a sum:
\begin{equation}
H\ =\ H_0+V
\end{equation}
of the highly symmetrical regular Hamiltonian $H_0$ (say, of non-interacting
particles or quasi-particles in the spherically-symmetrical mean field):
\begin{equation}
H_0\phi_k=\epsilon_k\phi_k
\end{equation}
and of the perturbation $V$ which destroys the symmetries of $H_0$ (the
pair-wise particle-particle forces in nuclear case). Expand now the eigenstates
$\psi_i$ of $H$ over the "regular" basis $\phi$:
\begin{equation}
\psi_i=\sum_k c^k_i\phi_k
\end{equation}
and look for the probability $P_k(E_i)=|c^k_i|^2$ to find the original
"regular" component $\phi_k$ in the different eigenstates $\psi_i$ (with
eigenenergies $E_i$) of our nonintegrable system. It is obvious, that for
sufficiently small perturbations $V$ the probability $P_k(E_i)$ is centered
around the "original" energy $\epsilon_k$ and tends to saturate to unity
over some characteristic energy interval $\Gamma_{spr}$ which is called
"the spreading width" of the initially unperturbed state $\phi_k$. Various
realistic models (see e.g. chapter 2 of ref.[6]) give the Lorentzian shape
of the strength function energy dependence:
\begin{equation}
S_k(E_i)=\frac{|c^k_i|^2}{D}\approx\frac{}{2\pi}
\frac{\Gamma_{spr}}{(E_i-\epsilon_k)^2+\Gamma_{spr}^2/4}
\end{equation}
where D is the average level spacing of the nonintegrable system. A slight
generalization of the derivation given in ref. [6] allows [3,4] to express the
spreading width in terms of the "mean square root" matrix element
$\tilde{v}=\sqrt{<v^2>}$ of the interaction $V$ mixing the basic states $\phi$
(angular brackets imply averaging over all the basic components admixed by
$V$ to a given one):
\begin{equation}
\Gamma_{spr}\sim \tilde{v}\sqrt{N_d}
\end{equation}
Here $N_d$ stands for the degeneracy rank of the initial level $\epsilon_k$.

Thus the system formally becomes nonintegrable as soon as $\Gamma_{spr}$
deviates from zero. However, while the ratio 
\begin{equation}
\ae=\frac{\Gamma_{spr}}{D_0}
\end{equation}
(where $D_0$ is the level spacing of the initial regular system) is smaller
than unity - the traces of the initial good quantum numbers are quite obvious
as isolated maxima of the strength function. We can easily distinguish
between the maxima corresponding to the different values of the originally
good quantum numbers.
This is the analogue of the classical "weak chaos" governed by the KAM
theorem.
When $\ae$ exceeds unity these traces of regularity disappear since it
becomes impossible to distinguish between the successive maxima of the
strength function corresponding to the different values of the original
quantum numbers $k$. This situation is the quantum analogue of the smearing
out and disappearance of the invariant tori. It means that we
approach the domain of "global" or "hard" chaos.

Furrier transforming Eq. (4) one can show (see e.g. [6]) that
$\Gamma_{spr}/\hbar$ defines the rate of decay of the "regular" states
$\phi$ resulting from the instability caused by the perturbation $V$. One
can even
form the wave packets $|A>$ of the states $\phi_k$ and analyze the recurrence
probabilities $P(t)=|<A(t)|A(0)>|^2$. This analysis shows [3] the periodic
recurrences with classical period $T$ modulated exponentially by the factor
$exp(-\Gamma_{spr}t/\hbar)$ arising from the above instability. Combining
these results with the results of Heller's wave-packet experience
(see e.g. [7] or paragraph 15.6 of ref. [8]),
one can show that the quantity $\Gamma_{spr}/\hbar$ transforms in the
classical limit into the Lyapunov exponent $\Lambda$:
\begin{equation}
\frac{\Gamma_{spr}}{\hbar}\rightarrow\Lambda
\end{equation}
The corresponding classical limit for the dimensionless chaoticity criterion
is:
\begin{equation}
\ae\rightarrow\frac{\Lambda T}{2\pi}=\frac{\chi}{2\pi}
\end{equation}
where $T$ is the classical period and $\chi$ is the stability parameter of
the classical monodromy matrix (see, e.g. [8]).

Thus the particular quantity $\Gamma_{spr}$ and the parameter $\ae$ seem to
be more accurate numerical
measures of quantum chaoticity than the level distribution law - this is
proved by the nuclear physics experience [2-4] and by its application
to one of the most popular in classical mechanics cases of transition
from regularity to chaos - Hennon-Heiles problem [5].

\section{Diamagnetic Kepler Problem}

Another very popular model for studies of transition from regularity
to chaos in classical mechanics is the non-relativistic hydrogen atom in
the uniform magnetic field (see e.g. [8,10,14]) with the Hamiltonian:
\begin{equation}
H=p^2/2m -e^2/r +\omega l_z  +\frac{1}{2}m\omega ^2(x^2+y^2)
\end{equation}
Here the frequency $\omega=eB/2mc$ is a half of the cyclotron frequency
and $B$ is the strength of
the magnetic field acting along z-axis. The dimensionless field strength
parameter $\gamma=\hbar\omega/{\cal R}$ (here ${\cal R}$ is the Rydberg energy) is
usually combined with the electron energy $E$ to produce the scaled energy
$\epsilon=E\gamma^{-2/3}$. When the scaled energy varies from $-\infty$
(for $B=0$) to 0 (for $B=\infty$) the regular motion of the system becomes
more and more chaotic. The fraction $R$ of available phase space covered by
regular trajectories was calculated in ref. [9 - 10] as a function of scaled
energy for the case of $l_z=0$ (see Fig.1), showing the rapid chaotisation of
the system in the range $-0.48\le\epsilon\le -0.125$.

We analyzed the quantum analogue of this system (with the Hamiltonian
(9))
on the same lines as it was done in [5] for the quantum Henon-Heiles problem,
namely we traced the gradual destruction of the $O(4)$ symmetry characteristic
of the unperturbed motion in Coulomb potential by the external magnetic field
B. In other words, we traced the disappearance of the "good" quantum numbers
(integrals of motion) which characterize the regular motion in this potential.
In order to do this, we diagonalized the Hamiltonian matrix (9) in parabolic
coordinates on the basis of purely Coulomb wave functions $\phi_{n_1 n_2 m}$,
whose eigenvalues in the unperturbed case are defined by the principal
quantum number n:
$$ n=n_1+n_2+|m|+1 $$
and are highly ($n^2$ times) degenerate. Diagonalizing the Hamiltonian matrix,
we defined the new
eigenvalues $E_i$ and the eigenstates $\psi_i$ in terms of the expansion
coefficients $c^k_i$ (see Eq. (3)).
As a next step, we plotted the energy distribution
of Eq. (4) for the coefficients' squares of the $n$-th shell
over the "new" eigenstates. Fig. 2 shows the examples of these
distributions for $n=10$, $m=0$ and the magnetic field $\gamma$ equal to
$4\cdot 10^{-4}$, $6\cdot 10^{-4}$, $8\cdot 10^{-4}$ and $12\cdot 10^{-4}$,
respectively. In order to increase the statistical accuracy, we performed
the averaging over all the components of the basis with the same $n$ value,
as it is usually done in nuclear physics and as it was done in the case
of quantum Henon-Heiles problem [5].
Assuming now that the shape of these distributions is approximately
Lorentzian, like in the case of the neutron strength function in nuclear
physics, we define $\Gamma_{spr}$ as the energy range around the maximum
where the sum of the squares of the coefficients $\sum_i|c^k_i|^2$
saturates to 0.5. Thus obtained values of $\Gamma_{spr}$ were divided
then by the level spacing $D_0$ between the adjacent maxima of the strength
function to give the desired parameter
$\ae$. The plot of this parameter versus the scaled energy $\epsilon$ is
given in Fig. 1.

We see that our parameter reaches the critical value of $\ae=1$ at the
critical scaled energy $\epsilon\approx -0.45$ in fairly good agreement
with the classical critical value $\epsilon\approx -0.48$ of refs. [9,10].
It is worthwhile to remind here that in the previous studies of the quantum
diamagnetic Kepler problem [11 - 14] the existence was pointed of the
approximately good quantum number $K$, corresponding to the eigenvalues
of the operator $\Sigma$ built as a combination of the Runge-Lenz vector
$A$:
\begin{equation}
\Sigma=4A^2 - 5A_z^2
\end{equation}
The eigenstates of this operator are obtained [12, 13] by prediagonalization
of the unperturbed Coulomb basis within a single manifold $n$ (which
physically corresponds to the values of our $\ae\leq 1$). The appreciable
$K$-mixing (disappearance of the integral of motion $\Sigma$) starts [13, 14]
when $\gamma^2 n^7\approx 16$. In our case of $n=10$ this corresponds to
the scaled energy $\epsilon\approx -0.45$.

\section{Conclusion}
Thus we confirmed once more the plausibility of the suggested approach to
quantum chaoticity, based on its connection with symmetry-breaking of the
regular motion which makes the number $M$ of the system's global integrals of
motion less than the number $N$ of its degrees of freedom. We had also
demonstrated that the spreading width $\Gamma_{spr}$ and the dimensionless
parameter $\ae$ might serve a good quantitative criterion of quantum
chaoticity. Likewise in the case of Henon-Heiles problem [4], the critical
scaled energy value $\epsilon_c$ when the parameter $\ae$ reaches unity
corresponds to the onset of "global" chaos on the classical phase portrait
for the diamagnetic Kepler problem. Here, however, the origin of the
approximate regularity of the perturbed system for $\epsilon\leq\epsilon_c$
is more evident. Although formally the external magnetic field makes the
system nonintegrable by reducing the number of global integrals of motion
to $M=2$ (energy and $l_z$), the third approximate integral of motion
($\Sigma$)
survives much longer making the system practically regular.

We should add in conclusion that the importance of studying the particular
example of hydrogen atom in the uniformed magnetic field was  stressed (see,
e.g. [14]) because it "is not an abstract model system but a real physical
system that can be and has been studied in the laboratory". These studies
were indeed started in 1986 (see [15]). One should point, however, that
atomic nucleus is also "not an abstract model", whose experimental and
theoretical studies are going on for more than half a century. As we had
already mentioned, Wigner developed his random matrix approach in order
to describe the experimentally observed properties of compound nuclear
resonances. Since those times nuclear physics accumulated a vast arsenal
of theoretical methods which allow Schr{\"o}dinger's equation to be
solved in some effective manner, even when the system is not integrable and
its behavior is chaotic by the criteria of level repulsion. As shown in [2-4],
most of
them are based on the smallness of the chaoticity parameter $\ae$, which seems
to be the most important small parameter of nuclear physics.

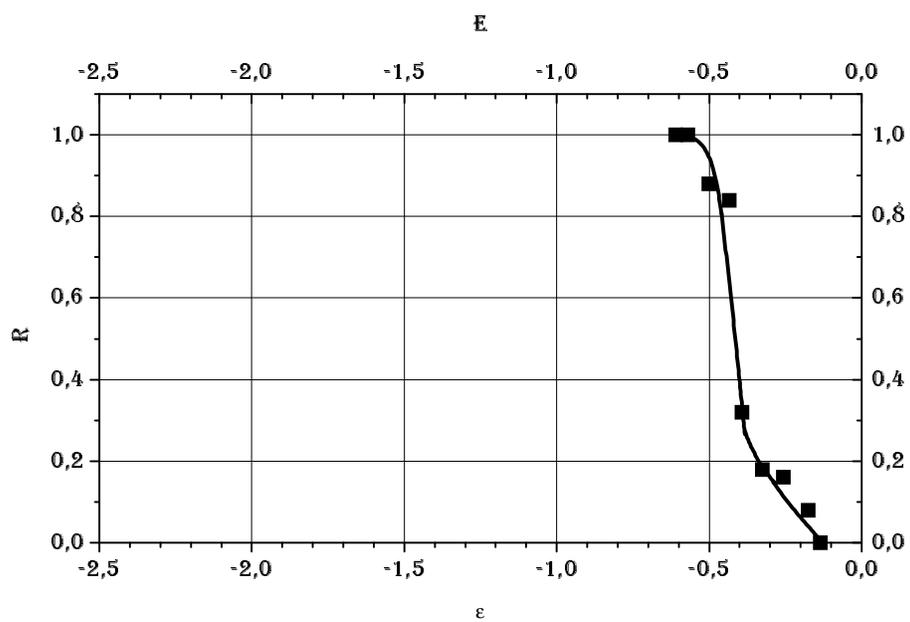
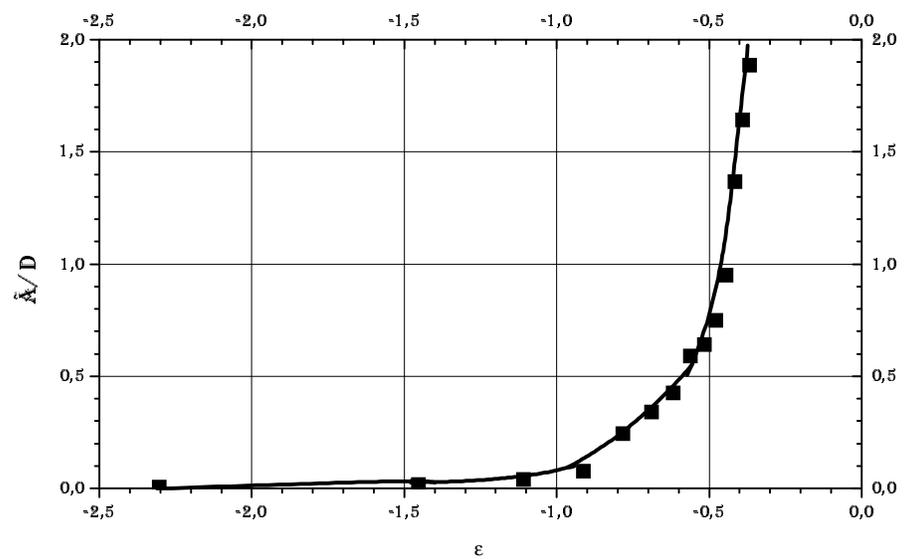
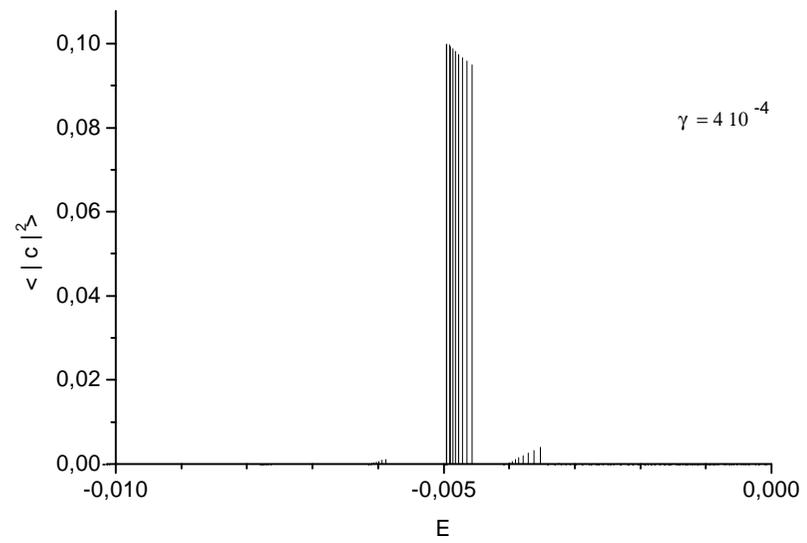
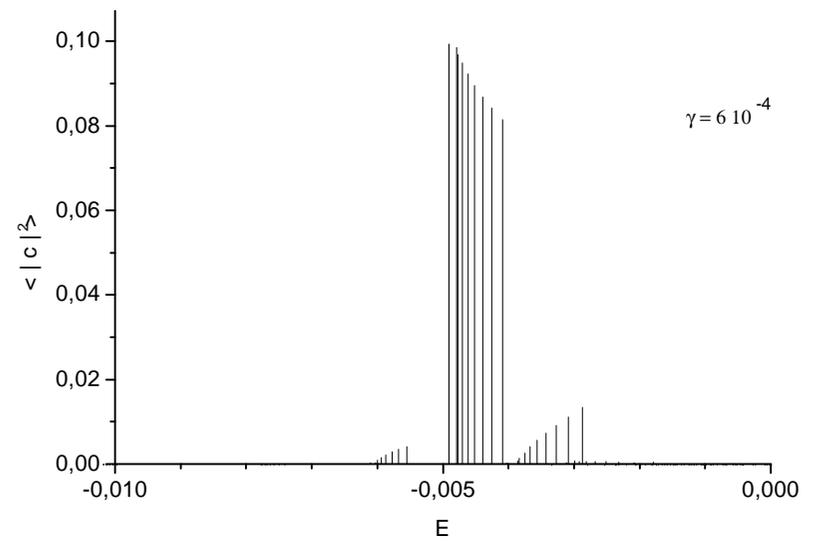
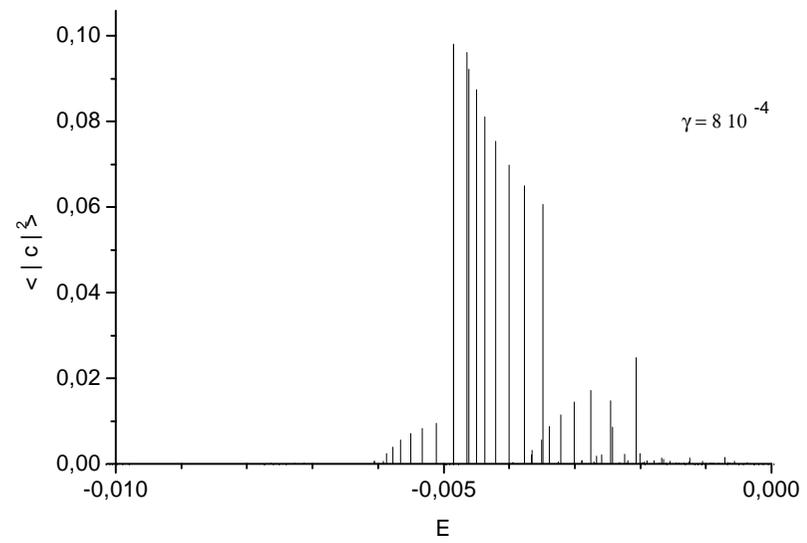
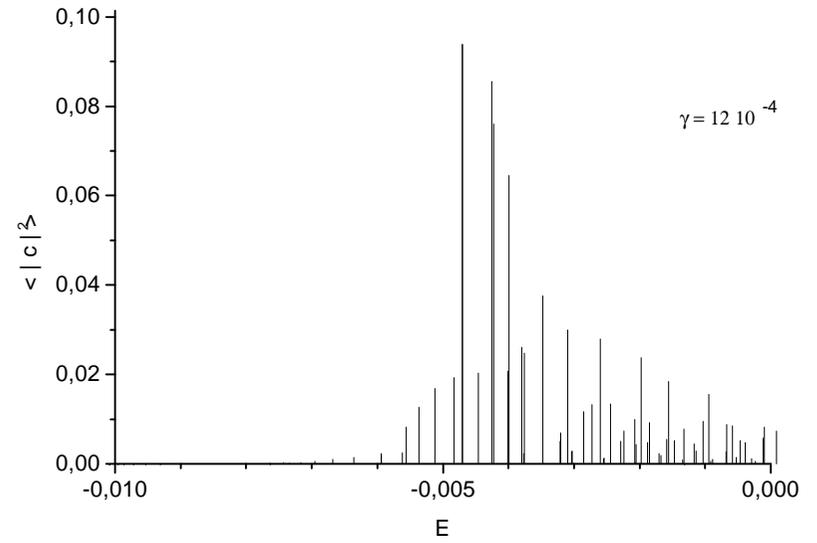
\end{document}